\documentclass[conference]{IEEEtran}
\IEEEoverridecommandlockouts

\usepackage{cite}
\usepackage{amsmath,amssymb,amsfonts}
\usepackage{algorithmic}
\usepackage{graphicx}
\usepackage{textcomp}
\usepackage{diagbox}
\usepackage{makecell}
\usepackage[table]{xcolor}
\usepackage{subcaption}
\usepackage{caption}
\def\BibTeX{{\rm B\kern-.05em{\sc i\kern-.025em b}\kern-.08em
    T\kern-.1667em\lower.7ex\hbox{E}\kern-.125emX}}

\begin{document}

\title{Iteration-Dependent Temporal Resilience in Diffusion Models for Approximate Computing}

\author{\IEEEauthorblockN{1\textsuperscript{st} Ori Schweitzer, 2\textsuperscript{nd} Uri Weiser}
\IEEEauthorblockA{\textit{The Faculty of Electrical and Computer Engineering} \\
\textit{Technion, Haifa, Israel}\\
ori.sch@campus.technion.ac.il, uri.weiser@ee.technion.ac.il}
\and

\IEEEauthorblockN{3\textsuperscript{rd} Freddy Gabbay}
\IEEEauthorblockA{\textit{The Institute of Electrical Engineering and Applied Physics} \\
\textit{The Hebrew University, Jerusalem, Israel}\\
freddy.gabbay@mail.huji.ac.il}
}

\maketitle
\begin{abstract}
This paper empirically analyzes iteration-dependent resilience in diffusion models by replacing MACs with efficient SAC approximations. Results on Q-Diffusion show early steps are highly sensitive while later steps are robust, enabling iteration-aware approximation that improves efficiency and reduces hardware cost with minimal quality loss.
\end{abstract}


\section{Introduction}
Generative AI models—especially diffusion models—achieve state-of-the-art image and video generation but require hundreds of sequential denoising iterations, each dominated by computation-intensive U-Net matrix operations \cite{10081412}. This results in high inference latency and substantial compute demands, limiting deployment on edge devices and high-throughput servers.

Deep neural networks (DNNs) often tolerate numerical noise, enabling approximate computing to reduce hardware area and power with minimal accuracy loss \cite{10813351,10123691,NEURIPS2021_9431c87f}. A common strategy replaces full-precision multiply–accumulate (MAC) units with low-power approximate computing alternatives. However, the temporal resilience of diffusion sampling—an inherently sequential process—remains poorly understood, as approximation errors may propagate differently across early, middle, and late denoising steps. This paper provides the first empirical characterization of the iteration-dependent temporal resilience of diffusion models to hardware-level approximation by replacing MACs with approximate computation across the sampling trajectory. As a case study, we employ a Shift-and-Accumulate (SAC) microarchitecture—an efficient MAC substitute based on our prior designs \cite{10123691}—that achieves over 30\% reduction in hardware area and power. We show that approximation sensitivity is highly non-uniform, with early iterations significantly more fragile than later ones.

Our contributions are threefold: (1) an empirical framework for measuring iteration-level resilience to SAC approximation; (2) the discovery of a strongly non-uniform resilience profile in which early steps are disproportionately sensitive; and (3) demonstration that this structure enables iteration-aware approximation strategies that improve computational efficiency while preserving image fidelity.

\subsection{Related Background}
\textbf{Diffusion Models.}
Diffusion models generate samples by iteratively transforming Gaussian noise into an image through a sequence of learned denoising steps. At each timestep $t$, a neural network $\epsilon_\theta(x_t, t)$ predicts the noise in the latent $x_t$, enabling the update toward $x_{t-1}$. Variants such as Denoising Diffusion Probabilistic Models (DDPM) and Denoising Diffusion Implicit Models (DDIM) differ in their sampling dynamics, yet all share the same multi-step temporal structure in which each iteration contributes differently to the denoising process. This sequential dependency makes diffusion models computationally intensive and motivates studying how approximation errors propagate across iterations.

\begin{figure}[htbp]
  \centering
  \includegraphics[width=\linewidth,height=0.22\textheight]{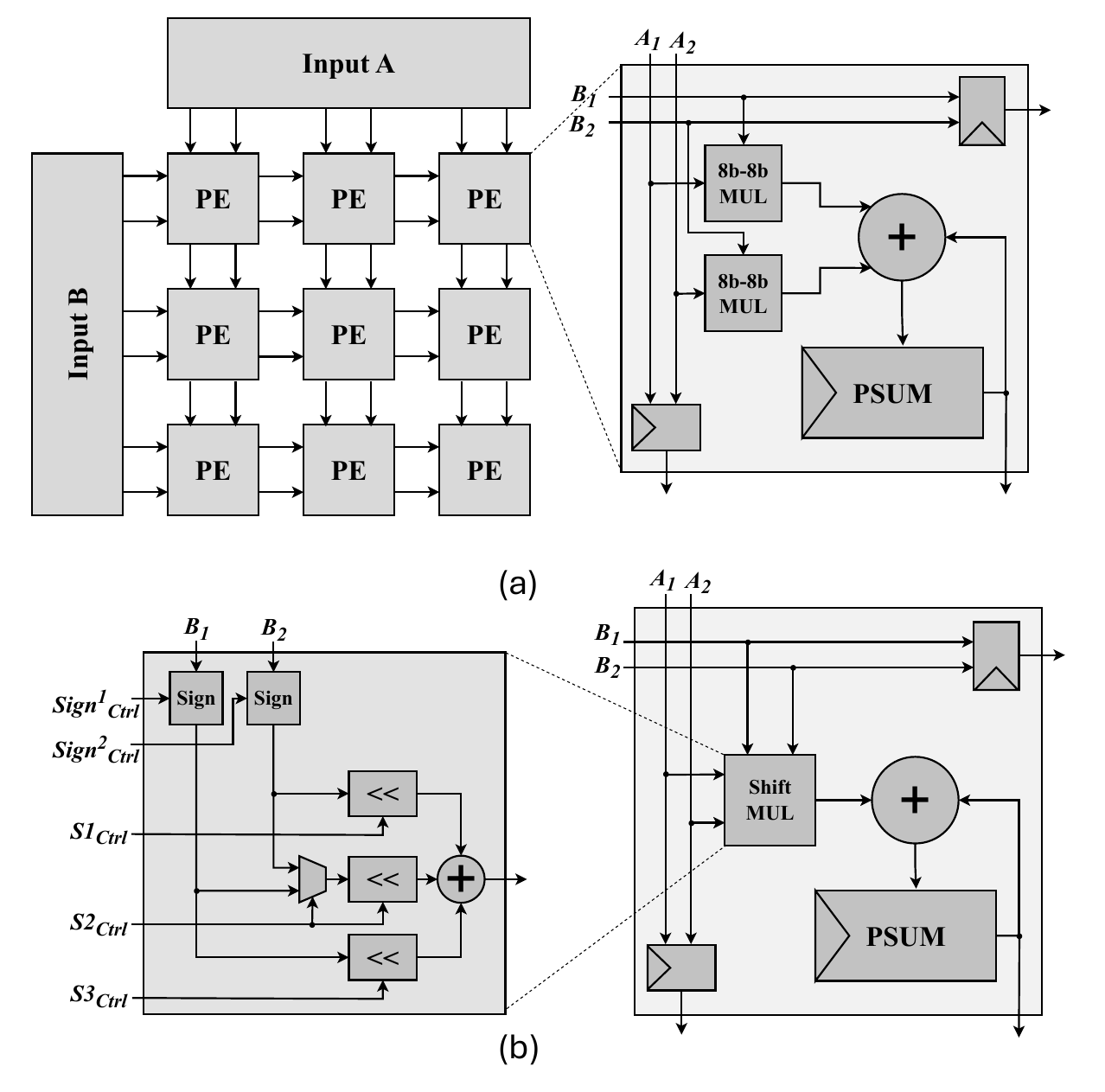}
  \caption{SMT SA using MAC and SAC units. (a) 3x3 SA using MAC-based, two-threaded PEs. (b) SAC-based, two-threaded PEs using three shifters, two allocated statically (one per thread) and one shared between both threads.}
  \label{fig:PE_SAC_MT}
\end{figure}

\textbf{SAC Systolic Array (SA) Microarchitecture.}
Unlike the conventional MAC-based design in Fig.~\ref{fig:PE_SAC_MT}(a), which uses full 8b--8b multipliers for each thread, the SA SAC-based architecture used in this paper as a case study \cite{10813351} (Fig.~\ref{fig:PE_SAC_MT}(b)) replaces multipliers with lightweight shift-and-accumulate logic. As in the baseline Simultaneous Multi-Threaded (SMT) design \cite{9251935}, each thread performs a single MAC operation; however, in SAC this MAC is \emph{approximated} by decomposing the first operand into a small set of leading-one positions and computing the product through corresponding shift-and-accumulate operations on the second operand. Each thread uses a \emph{static shifter} for its highest leading-one bit, while a shared \emph{dynamic shifter} serves the thread with the next-highest leading-one bit among active threads. The SAC microarchitecture reduces hardware complexity, achieving over 30\% area and power savings with under 1\% accuracy loss across CNNs, Vision Transformers, and LLMs \cite{10813351}.

\section{Temporal Resilience in Diffusion Models}
Diffusion sampling proceeds through a sequence of $T$ denoising iterations, each transforming a noisy latent $x_t$ toward the final image $x_0$. Because each timestep performs a distinct denoising operation, approximation errors introduced at different iterations may propagate and accumulate differently. To characterize this behavior, we empirically measure the effect of replacing exact MAC operations with SAC units at specific timesteps while keeping all other iterations exact. This procedure isolates the per-iteration sensitivity of the diffusion process to approximate arithmetic.

\begin{figure}[htbp]
  \centering
  \includegraphics[width=\linewidth,height=0.15\textheight]{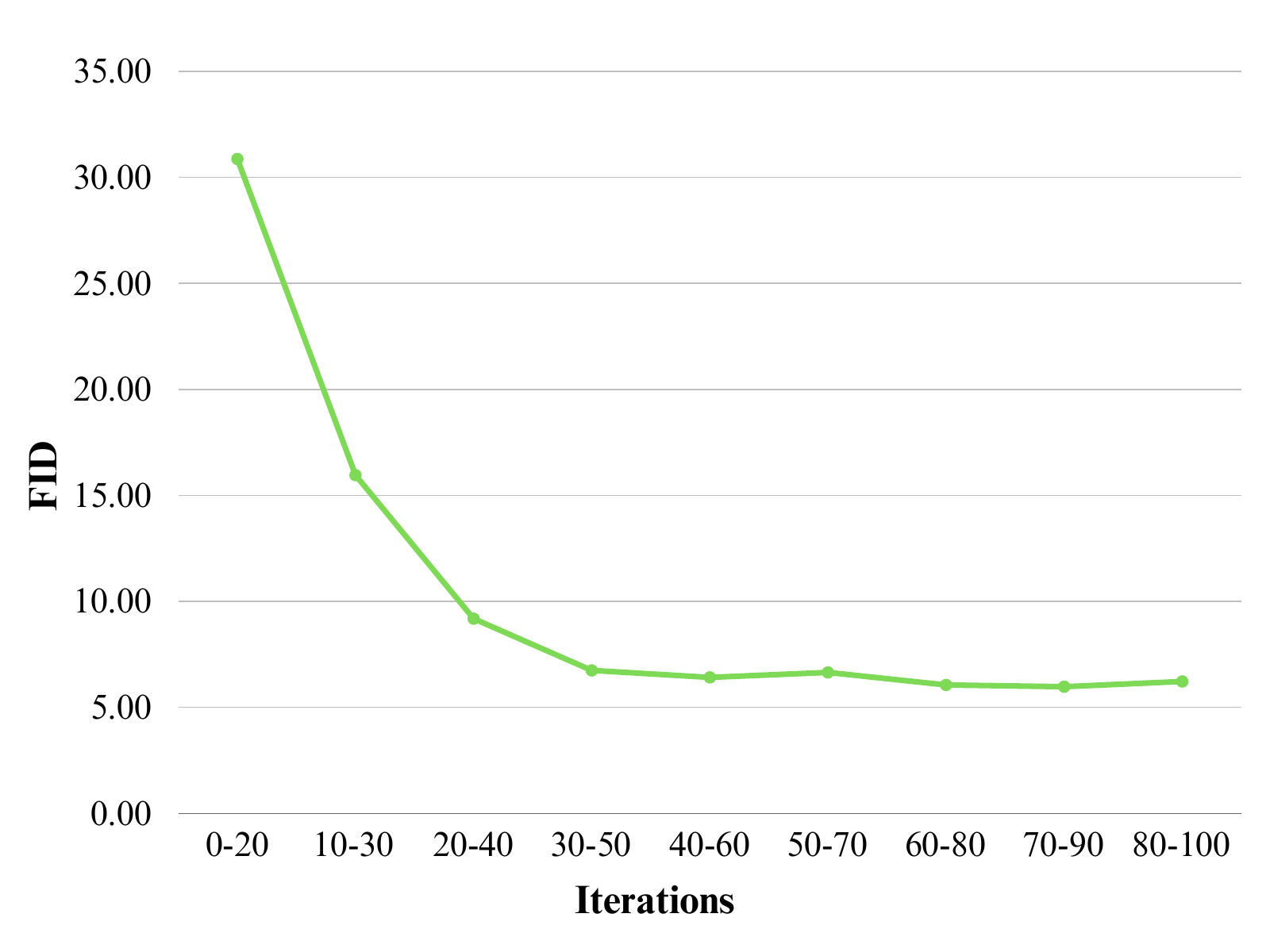}
  \caption{FID impact of applying SAC approximation at different sampling intervals in Q-Diffusion on CIFAR-10; other iterations use accurate MACs.}
  \label{fig:fid1}
\end{figure}

We evaluate temporal resilience using the 8-bit quantized Q-Diffusion model \cite{qdiffusion} on the CIFAR-10 \cite{cifar10} dataset. We measure degradation using the Fréchet Inception Distance (FID) \cite{yu2021frechet} and qualitative inspection.
Figure~\ref{fig:fid1} illustrates the impact of employing the SAC approximation at different parts of the Q-Diffusion sampling process on CIFAR-10, reported using FID. The model uses 100 DDIM iterations, and we apply a sliding window of 20 consecutive iterations in which SAC is activated, while all remaining iterations continue to use standard MAC operations. Each point on the graph corresponds to a different starting position of the sliding window.
 
 These results show that the diffusion model is more resilient in the later stages of the sampling process, while the first iterations are more sensitive to approximation. Starting from iteration 30, there is negligible impact of the approximated iteration on the FID scores. The sampled images produced with approximation at iterations 0 to 20 are much noisier and of lower quality, leading to worse FID scores. 

\begin{figure}[htbp]
  \centering
  \includegraphics[width=\linewidth]{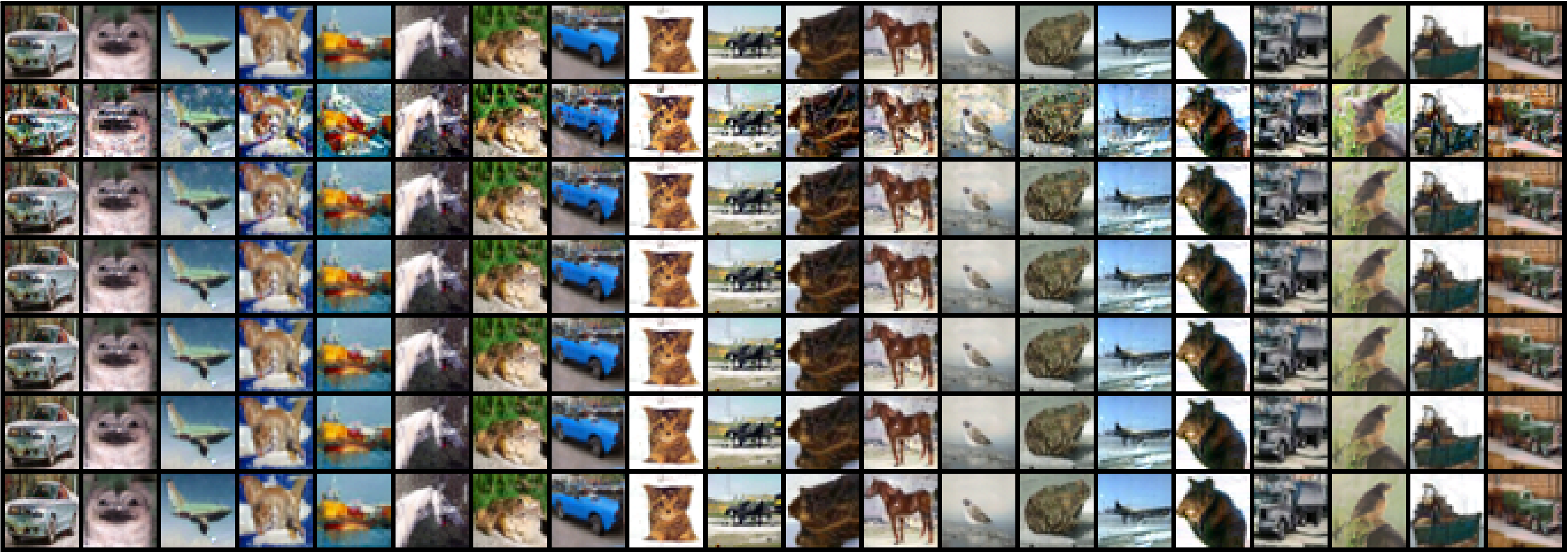}
  \caption{Approximation applied at different 20-step intervals in the Q-Diffusion sampling process on CIFAR-10. Top row: accurate MAC baseline. Subsequent rows show samples for SAC approximate windows ending at iteration 90, starting at iterations 10, 30, 40, 50, 60, and 70.}
  \label{fig:cifar}
\end{figure}

\begin{figure}[htbp]
  \centering
  \includegraphics[width=\linewidth,height=0.15\textheight]{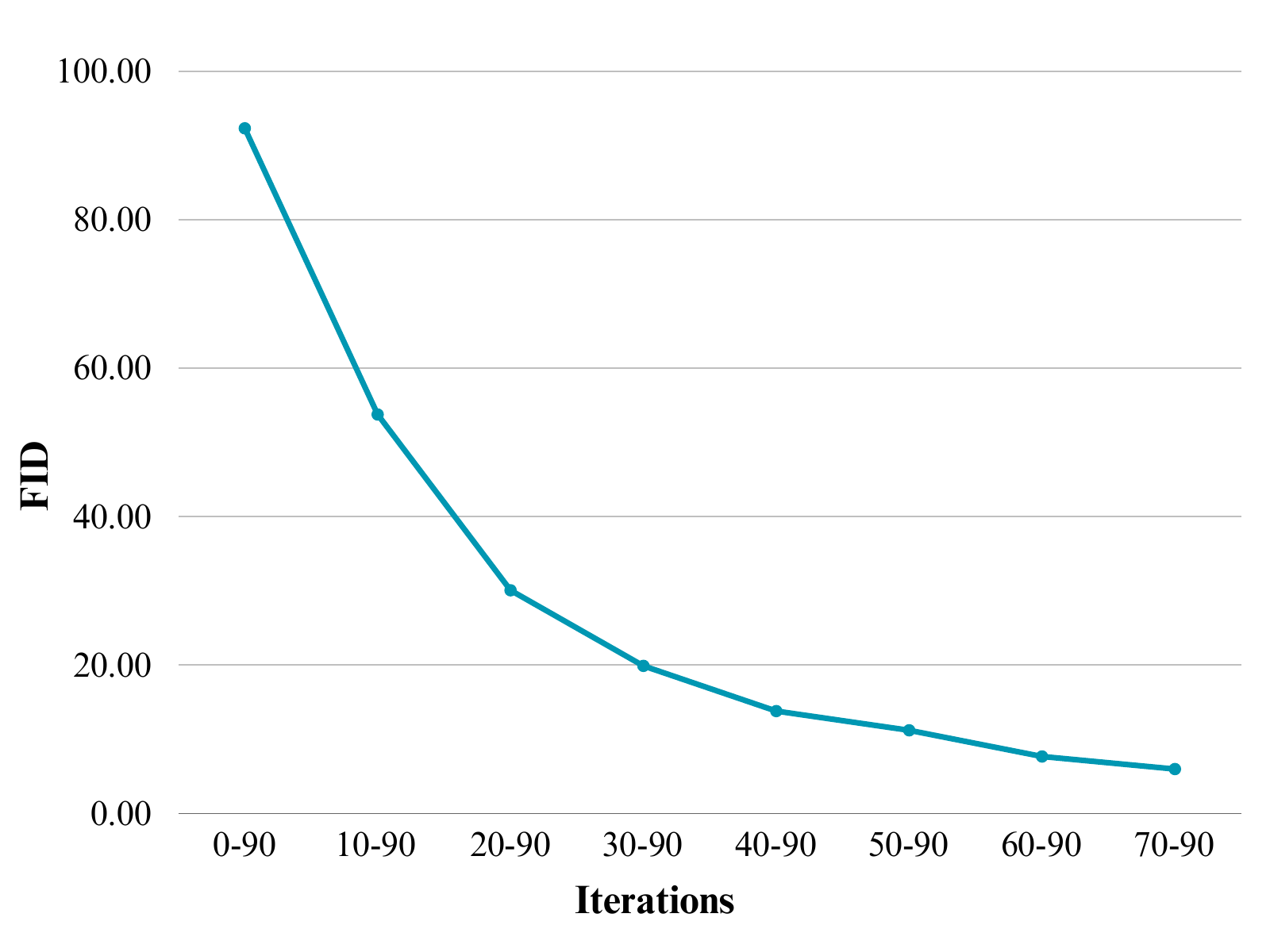}
  \caption{FID impact of applying SAC-based approximation at different sampling intervals in Q-Diffusion on CIFAR-10; all other iterations use accurate MAC operations.}
  \label{fig:fid2}
\end{figure}

To improve processing efficiency, we increase the number of sampling iterations executed with approximate SAC computations and evaluate the resulting impact on CIFAR-10, as shown in Figures~\ref{fig:cifar} and~\ref{fig:fid2}. Approximation is applied in the later stages of the sampling trajectory—where resilience is highest—while the final ten iterations remain accurate MAC computations to ensure fine-grained refinement and higher image fidelity. As expected, image quality improves as fewer iterations use approximation; however, even when half of the iterations (50 out of 100) employ SAC, the model still produces reasonable images, albeit with a noticeable degradation in FID.

Our findings show that diffusion sampling exhibits a structured temporal robustness that approximate accelerators can exploit. Because later iterations tolerate aggressive approximation, hardware can pipeline precise units for early sensitive steps with SAC-based approximate units for later robust steps. This enables substantial efficiency gains and reductions in area and power without compromising image fidelity.

\section{Conclusion}
This paper presented the first empirical study of iteration-dependent temporal resilience in diffusion models under hardware-level approximation. Using SAC as a case study, we selectively replaced MACs at different points in the sampling trajectory and showed that early denoising steps are highly sensitive to approximation, whereas later steps are substantially more robust. This insight enables iteration-aware approximation strategies that apply accurate computation to early iterations and efficient approximate processing to later ones, reducing computational cost, hardware area, and power with minimal impact on image quality.



\begin{thebibliography}{1}
\providecommand{\url}[1]{#1}
\csname url@samestyle\endcsname
\providecommand{\newblock}{\relax}
\providecommand{\bibinfo}[2]{#2}
\providecommand{\BIBentrySTDinterwordspacing}{\spaceskip=0pt\relax}
\providecommand{\BIBentryALTinterwordstretchfactor}{4}
\providecommand{\BIBentryALTinterwordspacing}{\spaceskip=\fontdimen2\font plus
\BIBentryALTinterwordstretchfactor\fontdimen3\font minus \fontdimen4\font\relax}
\providecommand{\BIBforeignlanguage}[2]{{%
\expandafter\ifx\csname l@#1\endcsname\relax
\typeout{** WARNING: IEEEtran.bst: No hyphenation pattern has been}%
\typeout{** loaded for the language `#1'. Using the pattern for}%
\typeout{** the default language instead.}%
\else
\language=\csname l@#1\endcsname
\fi
#2}}
\providecommand{\BIBdecl}{\relax}
\BIBdecl

\bibitem{10081412}
F.-A. Croitoru, V.~Hondru, R.~T. Ionescu, and M.~Shah, ``Diffusion models in vision: A survey,'' \emph{IEEE Transactions on Pattern Analysis and Machine Intelligence}, vol.~45, no.~9, pp. 10\,850--10\,869, 2023.

\bibitem{10813351}
O.~Schweitzer, U.~Weiser, and F.~Gabbay, ``Enhancing dnn computational efficiency via decomposition and approximation,'' \emph{IEEE Access}, vol.~12, pp. 197\,347--197\,362, 2024.

\bibitem{10123691}
S.~Kurzum, G.~Shomron, F.~Gabbay, and U.~Weiser, ``Enhancing dnn training efficiency via dynamic asymmetric architecture,'' \emph{IEEE Computer Architecture Letters}, vol.~22, no.~1, pp. 49--52, 2023.

\bibitem{NEURIPS2021_9431c87f}
G.~Shomron, F.~Gabbay, S.~Kurzum, and U.~Weiser, ``Post-training sparsity-aware quantization,'' in \emph{Advances in Neural Information Processing Systems}, vol.~34.\hskip 1em plus 0.5em minus 0.4em\relax Curran Associates, Inc., 2021, pp. 17\,737--17\,748.

\bibitem{9251935}
G.~Shomron and U.~Weiser, ``Non-blocking simultaneous multithreading: Embracing the resiliency of deep neural networks,'' in \emph{2020 53rd Annual IEEE/ACM International Symposium on Microarchitecture (MICRO)}, 2020, pp. 256--269.

\bibitem{qdiffusion}
X.~Li, Y.~Liu, L.~Lian, H.~Yang, Z.~Dong, D.~Kang, S.~Zhang, and K.~Keutzer, ``Q-diffusion: Quantizing diffusion models,'' in \emph{Proceedings of the IEEE/CVF International Conference on Computer Vision}, 2023, pp. 17\,535--17\,545.

\bibitem{cifar10}
A.~Krizhevsky, G.~Hinton \emph{et~al.}, ``Learning multiple layers of features from tiny images,'' 2009.

\bibitem{yu2021frechet}
Y.~Yu, W.~Zhang, and Y.~Deng, ``Frechet inception distance (fid) for evaluating gans,'' \emph{China University of Mining Technology Beijing Graduate School}, vol.~3, no.~11, 2021.

\end{thebibliography}
\end{document}